\documentclass[letterpaper, 10pt]{article}
\title{Applications of Multi-Valued Quantum Algorithms}
\author{Yale Fan}
\date{April 20, 2007\footnote{updated July 4, 2010}}
\input{Qcircuit}
\input xy
\xyoption{all}
\usepackage{amsmath}
\newtheorem{QFT}{Definition}
\newtheorem{function}[QFT]{Definition}
\newtheorem{unity}{Lemma}
\newtheorem{affine}{Theorem}
\newtheorem{D-J}[affine]{Theorem}
\newtheorem{fourth}[affine]{Theorem}
\begin{document}
\maketitle
\begin{abstract}
This paper generalizes both the binary Deutsch-Jozsa and Grover algorithms to $n$-valued logic using the quantum Fourier transform.  Our extended Deutsch-Jozsa algorithm is not only able to distinguish between constant and balanced Boolean functions in a single query, but can also find closed expressions for classes of affine logical functions in quantum oracles, accurate to a constant term.  Furthermore, our multi-valued extension of the Grover algorithm for quantum database search requires fewer qudits and hence a substantially smaller memory register, as well as fewer wasted information states, to implement.  We note several applications of these algorithms and their advantages over the binary cases.
\end{abstract}

\section{Introduction}

Quantum computers have the ability to implement algorithms in inherently multi-valued logic.  Multi-valued quantum logic gates have already been developed by Muthukrishnan and Stroud \cite{MS}, and shown to be feasible in a linear ion trap model.  Generalizing quantum algorithms to multi-valued logic can also present marked advantages over other binary classical and quantum algorithms.
\vspace{5 mm}

The original binary Deutsch-Jozsa algorithm \cite{Deutsch} considers a Boolean function of the form $f:\{0,1\}^r \rightarrow \{0,1\}$ implemented in a black box circuit, or oracle, $U_f$.  Input states are put in a quantum superposition as query ($x$) and answer ($y$) registers so that their state vectors are expressed in terms of the dual basis \cite{Jozsa}
\[
|0'\rangle = \frac{1}{\sqrt{2}}(|0\rangle + |1\rangle) \mbox{ and } |1'\rangle = \frac{1}{\sqrt{2}}(|0\rangle - |1\rangle) \mbox{, also denoted } |+\rangle \mbox{ and } |-\rangle.
\]
The oracle is defined by its action on the registers: $U_f|xy\rangle = |x\rangle|y \oplus f(x)\rangle$, where the $|x\rangle$ register is the tensor product of input states $|x_1\rangle\cdots|x_r\rangle$.  When it is promised that the function in question is either constant (returning a fixed value) or balanced (returning outputs equally among 0 and 1), the algorithm decides deterministically which type it is with a single oracle query as opposed to the $2^{r-1}+1$ required classically.  The corresponding circuit is shown in Figure 1 ($/^r$ denotes $r$ wires in parallel).

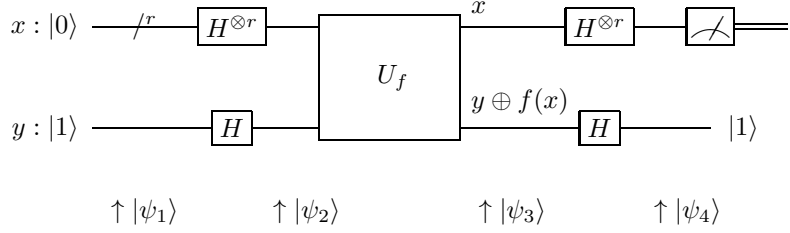
\begin{figure}[ht]
\hspace{.5 cm}\centerline{ \Qcircuit @C=2em @R=2.5em { & \lstick{x: \ket{0}} & {/^r} \qw & \gate{H^{\otimes r}} & \multigate{1}{\hspace{.6 cm} U_f \hspace{.5 cm}} & \ustick{x\hspace{.9 cm}} \qw & \gate{H^{\otimes r}} & \meter & \cw \\
& \lstick{y: \ket{1}} & \qw & \gate{H} & \ghost{\hspace{.6 cm} U_f
\hspace{.5 cm}} & \ustick{\hspace{.2 cm} y\oplus f(x)} \qw &
\gate{H} & \qw & & \lstick{\ket{1}\hspace{.9 cm}}\\
& & \uparrow|\psi_1\rangle & \hspace{2 cm} \uparrow|\psi_2\rangle &
& \uparrow|\psi_3\rangle & & \uparrow|\psi_4\rangle \hspace{.6 cm}}}
\vspace{1 mm} \caption{The Deutsch-Jozsa circuit}
\end{figure}

Grover's algorithm \cite{Grover} is a probabilistic search algorithm usually presented in the context of searching an unsorted database.  It consists of the iteration of a compound ``Grover operator,'' which consists of an oracle and an inversion-about-the-average operator $D$, on a superposed register of search states as well as an ancillary qubit.  The oracle marks the searched state by rotating its phase by $\pi$, while the inversion-about-the-average operator increases the amplitude of this state.
\vspace{5 mm}

In this paper, we present extensions of both the Deutsch-Jozsa and Grover algorithms to arbitrary radices of multi-valued quantum logic.  We denote addition over the additive group $\mathbf{Z}_n$ by the symbol $\oplus$ and the Kronecker tensor product by $\otimes$.

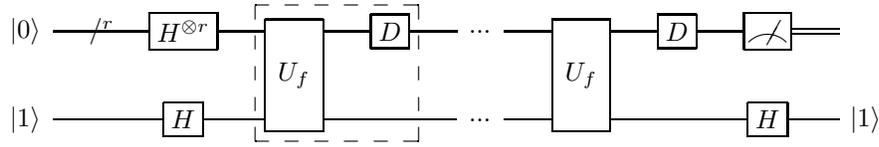
\begin{figure}[ht]
\hspace{.5 cm}\centerline{ \Qcircuit @C=1.8em @R=2em { & & & & \mbox{G Operator \hspace{-1.2 cm}} & & & & & & & & \\
& \lstick{\ket{0}} & {/^r} \qw & \gate{H^{\otimes r}} & \multigate{1}{U_f} & \gate{D} & \qw & \lstick{...} & \multigate{1}{U_f} & \gate{D} & \meter & \cw \\
& \lstick{\ket{1}} & \qw & \gate{H} & \ghost{U_f} & \qw & \qw & \lstick{...} & \ghost{U_f} & \qw & \gate{H} & \qw & \lstick{\ket{1}\hspace{-.1 cm}} \gategroup{2}{5}{3}{6}{.7 em}{--}}}
\vspace{1 mm} \caption{The Grover circuit}
\end{figure}

The Hadamard transform is a special case of the quantum Fourier transform (QFT) in Hilbert space $\mathcal{H}_n$.  The well-known Chrestenson gate for ternary quantum computing is also equivalent to the Fourier transform in $\mathcal{H}_3$.  Cereceda \cite{Cereceda} has generalized the Deutsch algorithm using two qudits to $d$-dimensional quantum systems where $d=2^k$.  However, placing no restrictions on the number of computational basis states $n$ leads to a far more versatile characterization of the Deutsch-Jozsa algorithm.  The Fourier matrix of order $n$ over the $n^{\mathrm{th}}$ roots of unity $\omega^k = e^{i2\pi k/n}$ is as follows:

\[
{\cal{F}}_n = \frac{1}{\sqrt{n}}\left( \begin{array}{ccccccccc}
1 & 1 & 1 & \cdots & 1 \\
1 & \omega & \omega^2 & \cdots & \omega^{n-1} \\
1 & \omega^2 & \omega^4 & \cdots & \omega^{2(n-1)} \\
1 & \omega^3 & \omega^6 & \cdots & \omega^{3(n-1)} \\
\vdots & \vdots & \vdots & \ddots & \vdots \\
1 & \omega^{n-1} & \omega^{2(n-1)} & \cdots & \omega^{(n-1)(n-1)}
\end{array} \right).
\]

\begin{QFT}
The action of the quantum Fourier transform is described by $\displaystyle \mbox{QFT}_n : |j\rangle \mapsto \frac{1}{\sqrt{n}}\sum_{k=0}^{n-1}e^{i2\pi jk/n} |k\rangle$ for $j \in \mathbf{Z}_n$ \emph{\cite{Jozsa}}.  The QFT matrix can be expressed as
\[
{\cal{F}}_n = \frac{1}{\sqrt{n}}\sum_{j=0}^{n-1}\sum_{k=0}^{n-1}e^{i2\pi jk/n} |j\rangle\langle k|.
\]
\end{QFT}

\section{The Deutsch-Jozsa Algorithm in Radix \emph{n}}

We will use the QFT in place of the Hadamard transform in the multi-valued equivalent of the Deutsch-Jozsa algorithm.  First, we define the functions that we are working with.

\begin{function}
An $r$-qudit multi-valued function of the form
\[
f:\{0, 1, \ldots, n - 1\}^r \rightarrow \{0, 1, \ldots, n - 1\}
\]
is constant when $f(x) = f(y) \mbox{ } \forall x, y \in \{0, 1, \ldots, n - 1\}^r$ and is balanced when an equal number of the $n^r$ domain values, namely $n^{r - 1}$, is mapped to each of the $n$ elements in the codomain.
\end{function}

In multi-valued logic, there are $n$ constant functions mapping each element in $\mathbf{Z}_n$ to a fixed element and $n!$ balanced permutative (bijective) mappings of single-qudit inputs.  For functions on $r$ qudits, there are accordingly $n^r!/n^{r - 1}!^n$ balanced (surjective) mappings.

\begin{affine}
All affine functions defined as $f(x_1, \ldots, x_r) = A_0 \oplus A_1x_1 \oplus \cdots \oplus A_rx_r$ with $A_0, \ldots, A_r \in \mathbf{Z}_n$ are either constant or balanced functions of $r$ qudits.
\end{affine}
\emph{Proof.} Those affine functions for which all coefficients $A_{i\neq0} = 0$ are constant.  For affine functions with at least one nonzero coefficient of $x_i$, the domain is $\{0, 1, \ldots, n-1\}^r$, which can be identified with $\{0, 1, \ldots, n^r - 1\} = \bigcup_m S_m$ where $S_m = \{m, n + m, 2n + m, \ldots, n^r - n + m\}$ for every $m \in \{0, 1, \ldots, n - 1\}$.  Each $S_m$ is a set of size $n^{r - 1}$, each of whose elements reduces modulo $n$ to $m$.  Since $f(p) = f(q)$ if $p\equiv q \mbox{ (mod }n)$, every element in the codomain $\{0, 1, \ldots, n - 1\}$ is assigned to exactly $n^{r - 1}$ different elements in the domain.  Such affine functions
satisfy the definition of a balanced function.
\vspace{5 mm}

\noindent The proof of the $n$-ary Deutsch-Jozsa algorithm will be aided by a simple lemma:

\begin{unity}
Primitive $n^\mathrm{th}$ roots of unity $\omega$ satisfy $\displaystyle \sum_{k=0}^{n-1}\omega^{\alpha k} = 0$ for integers $\alpha \not\equiv 0\pmod{n}$.
\end{unity}
\emph{Proof.} Consider the polynomial $z^n - 1 = 0$, of which $\omega^\alpha$ is a root.  1 being a real root for all integers $n$, this can be factorized as $(z - 1)(z^{n - 1} + z^{n - 2} + \cdots + 1) = 0$.  Since $\alpha \equiv r\pmod{n}$ for some $r$ where $0 < r < n-1$, $\omega^\alpha = \omega^r \neq 1$.  Therefore, $\sum_{k=0}^{n-1}z^k = 0$ for $z = \omega^\alpha$.
\vspace{5 mm}

\noindent This leads to our main result.

\begin{D-J}
The $n$-ary Deutsch-Jozsa algorithm applied to multi-valued functions of $r$ qudits can both distinguish between constant and balanced functions with a single oracle query and determine a closed expression for an affine function in $U_f$, excepting the constant term, as follows:

1. The constant term $A_0$ is preserved in the phase of the $x$-register at output $(\omega^{-A_0})$, which is lost during measurement.

2. The coefficients $A_1, \ldots, A_r$ are determined by the state of the $x$-register at output, $|A_1, \ldots, A_r\rangle$.
\end{D-J}

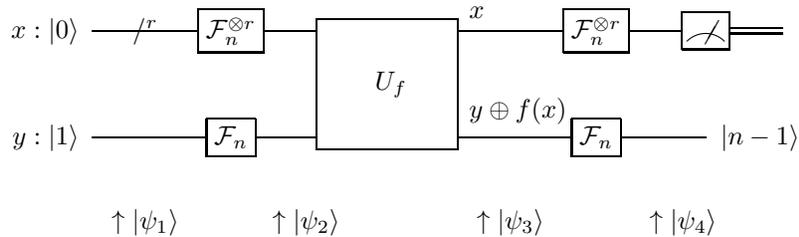
\begin{figure}[ht]
\hspace{.5 cm}\centerline{ \Qcircuit @C=2em @R=2.5em { & \lstick{x: \ket{0}} & {/^r} \qw & \gate{{\cal{F}}_n^{\otimes r}} & \multigate{1}{\hspace{.6 cm} U_f \hspace{.5 cm}} & \ustick{x\hspace{.9 cm}} \qw & \gate{{\cal{F}}_n^{\otimes r}} & \meter & \cw \\
& \lstick{y: \ket{1}} & \qw & \gate{{\cal{F}}_n} & \ghost{\hspace{.6
cm} U_f \hspace{.5 cm}} & \ustick{\hspace{.2 cm} y\oplus f(x)} \qw &
\gate{{\cal{F}}_n} & \qw & & \lstick{\ket{n-1}\hspace{.3 cm}}\\
& & \uparrow|\psi_1\rangle & \hspace{2 cm} \uparrow|\psi_2\rangle &
& \uparrow|\psi_3\rangle & & \uparrow|\psi_4\rangle \hspace{.6 cm}}}
\vspace{1 mm} \caption{The $n$-ary Deutsch-Jozsa circuit}
\end{figure}

In practice, the $y$-register would not be measured, but we follow through with the relevant calculations to demonstrate that its state at the output is constant, regardless of the function in the oracle.  The $x$- and $y$-registers are written separately as factors of the entire tensored state of the circuit at each step $|\psi_i\rangle$.
\vspace{5 mm}

First, we consider the case in which the function $f(x)$ hidden in the oracle is constant:
\[
|\psi_1\rangle=|0\rangle^{\otimes r}|1\rangle
\]\[
\xrightarrow{{\cal{F}}_n^{\otimes r + 1}}|\psi_2\rangle=\frac{1}{\sqrt{n^r}}\sum_{x=0}^{{n^r}-1}|x\rangle\otimes\frac{1}{\sqrt{n}}\sum_{y=0}^{n-1}e^{i2\pi y/n}|y\rangle
\]\[
\stackrel{U_f}{\longrightarrow} |\psi_3\rangle = \frac{1}{\sqrt{n^r}}\sum_{x=0}^{{n^r}-1}|x\rangle\otimes\frac{1}{\sqrt{n}}\sum_{y=0}^{n-1}e^{i2\pi y/n}|y\oplus f(x)\rangle.
\]
At this point, we can transfer the action of $U_f$ from the basis states themselves onto their phases by observing that if a basis vector $|j\oplus k\rangle$ is appended with the phase factor $\phi^{j}$, then $|j\rangle$ itself must have phase $\phi^{j-k}$.  This yields:
\begin{eqnarray*}
|\psi_3\rangle &=& \frac{1}{\sqrt{n^r}}\sum_{x=0}^{{n^r}-1}|x\rangle\otimes\frac{1}{\sqrt{n}}\sum_{y=0}^{n-1}e^{i2\pi[y-f(x)]/n}|y\rangle \\
&=& \frac{1}{\sqrt{n^r}}e^{-i2\pi f(x)/n}\sum_{x=0}^{{n^r}-1}|x\rangle\otimes\frac{1}{\sqrt{n}}\sum_{y=0}^{n-1}e^{i2\pi y/n}|y\rangle.
\end{eqnarray*}
Because we assume our function to be constant, $e^{-i2\pi f(x)/n}$ can be regarded as a global phase factor.  Subsequently, the QFT on the $x$-register can be computed explicitly:
\[
{{\cal{F}}_n}^{\otimes r}=\frac{1}{\sqrt{n}}\sum_{j=0}^{{n^r}-1}\sum_{k=0}^{{n^r}-1}e^{i2\pi jk/n}|j\rangle\langle k| \mbox{, giving}
\]\[
|\psi_3\rangle \xrightarrow{{\cal{F}}_n^{\otimes r + 1}}|\psi_4\rangle =
\]\[
\frac{1}{n^r}e^{-i2\pi f(x)/n}\sum_{j=0}^{{n^r}-1}\sum_{k=0}^{{n^r}-1}\sum_{x=0}^{{n^r}-1}e^{i2\pi jk/n}|j\rangle\langle k|x\rangle \hspace{3 pt}\otimes \mbox{ } \frac{1}{n} \sum_{j=0}^{n-1} \sum_{k=0}^{n-1} \sum_{y=0}^{n-1}e^{i2\pi (jk\oplus y)/n}|j\rangle\langle k|y\rangle.
\]
In the standard basis, the inner product $\langle k|z\rangle=\delta_{kz}$.  We can hence reduce the above to
\[
|\psi_4\rangle = \frac{1}{n^r}e^{-i2\pi f(x)/n}\sum_{j=0}^{{n^r}-1}\sum_{k=0}^{{n^r}-1}e^{i2\pi jk/n} |j\rangle \hspace{3 pt} \otimes \hspace{3 pt} \frac{1}{n}\sum_{j=0}^{n-1}\sum_{k=0}^{n-1}e^{i2\pi (j\oplus 1)k/n}|j\rangle.
\]
By Lemma 1, all basis states $|j\rangle$ in the $x$-register will have null amplitudes for $j\neq 0$.  Similarly, all basis states $|j\rangle$ in the $y$-register will have null amplitudes for $j\neq n-1$.  It follows that
\[
|\psi_4\rangle = |0\rangle^{\otimes r}|n-1\rangle
\]
with a phase factor of $e^{-i2\pi f(x)/n}$ for all constant functions $f(x)$.
\vspace{5 mm}

The balanced case is similar.  After initializing and superposing our states as above, we obtain:
\[
|\psi_3\rangle = \frac{1}{\sqrt{n^r}}\sum_{x=0}^{{n^r}-1}e^{-i2\pi f(x)/n}|x\rangle\otimes\frac{1}{\sqrt{n}}\sum_{y=0}^{n-1}e^{i2\pi y/n}|y\rangle.
\]
In this case, the phase factor $e^{-i2\pi f(x)/n}$ cannot be assumed to be global because its value is dependent upon $x$.  The output of the $y$-register will be the same as in the constant case, so we need only proceed with the state of the $x$-register.  After applying the second QFT:
\begin{eqnarray*}
|\psi_3\rangle\xrightarrow{{\cal{F}}_n^{\otimes r}}|\psi_4\rangle &=& \frac{1}{n^r}\sum_{j=0}^{{n^r}-1}\sum_{k=0}^{{n^r}-1}\sum_{x=0}^{{n^r}-1}e^{i2\pi jk/n}e^{-i2\pi f(x)/n}|j\rangle\langle k|x\rangle \\
&=& \frac{1}{n^r}\sum_{j=0}^{{n^r}-1}\sum_{x=0}^{{n^r}-1}e^{i2\pi [jx-f(x)]/n}|j\rangle.
\end{eqnarray*}
It is now necessary to show that $jx-f(x)=$ some constant $C$, or $f(x)=jx-C$, for a fixed value of $j\neq0$ and all $x$ in the domain $\{0,1,\ldots,{n^r}-1\}$.  This would allow $e^{i2\pi C/n}$, or $\omega^C$, to be the phase factor of some basis state $|j\rangle$ other than $|0\rangle^{\otimes r}$ as in the constant case, with a unit probability of measurement.  Equivalently, since addition and multiplication are modular, $f(x_1, \ldots, x_r)=-C\oplus j_1 x_1\oplus\cdots\oplus j_r x_r$ must hold for some $j\in \{1,...,{n^r}-1\}$, $j_i$ representing the $i^{\mathrm{th}}$ digit of $j$.  This is the definition of a non-constant affine function, so $j$ exists only when the hidden function in the oracle is affine.  The $x$-register is therefore measured to be:
\[
|\psi_4\rangle=e^{i2\pi [jx-f(x)]/n}|j\rangle=\omega^C|j\rangle,
\]
$f(x)$ being balanced for $j\neq0$ and constant otherwise.  In consequence, the Deutsch-Jozsa algorithm gives a deterministic output only when $f(x)$ is restricted to being either constant or balanced, \emph{and} affine (we will give an example of the algorithm applied to a non-affine function below).  However, observe that the state $|0\rangle^{\otimes r}$ will have a zero probability of measurement for all balanced functions, whether affine or not, because its amplitude is
\[
\frac{1}{n^r}\sum_{x=0}^{{n^r}-1}e^{i2\pi f(x)/n} = \frac{n^{r-1}}{n^r}\sum_{x=0}^{n-1}e^{i2\pi f(x)/n} = \frac{1}{n}\sum_{x=0}^{n-1}e^{i2\pi x/n} = 0
\]
by Definition 2 and Lemma 1.  Thus, the algorithm is still deterministic in the sense that it can always distinguish between constant and either affine \emph{or} non-affine balanced functions, although with no fixed output in the latter case.
\vspace{5 mm}

Finally, the generalized Deutsch-Jozsa algorithm has a useful property that becomes apparent in multi-valued logic; it can not only distinguish between constant and balanced functions, but can determine explicitly the function $f(x_1, \ldots, x_r)=A_0\oplus A_1 x_1\oplus \cdots \oplus A_r x_r$ implemented by the oracle excepting the constant term $A_0$, given that it is affine.  As calculated above, the constant term $A_0$ of such a function is encoded in the phase of the $x$-register at output ($A_0=-C$, where the phase factor is $\omega^C$), while the respective coefficients $A_1, \ldots, A_r$ of $x$ are determined by the basis vector $|j\rangle = |A_1, \ldots, A_r\rangle$.  Since the phase of the $x$-register is lost at measurement, only $A_0$ cannot be retrieved.  We regard affine functions that differ only in the constant term as a ``class.''
\vspace{5 mm}

The $y$-register can be ignored for ease of theoretical computation (not in practice) if the oracle, corresponding to the diagonal operator
\[
U_f = \sum_{x=0}^{{n^r}-1} e^{-i2\pi f(x)/n} |x\rangle\langle x|,
\]
directly encodes the action of $f(x)$ into the phase of the $x$-register.  We use this scheme below.
\vspace{5 mm}

\noindent \textbf{Example 1} (Deutsch-Jozsa for an affine function)

$U_f$ contains the following balanced function defined on two qutrits $|AB\rangle$:

\vspace{.1 cm} \begin{center} \setlength{\unitlength}{.5 cm}
\begin{picture}(4,4)
\put(1,3){\line(-1,1){.8}} \put(1,0){\line(0,1){3}}
\put(2,0){\line(0,1){3}} \put(3,0){\line(0,1){3}}
\put(4,0){\line(0,1){3}} \put(1,0){\line(1,0){3}}
\put(1,1){\line(1,0){3}} \put(1,2){\line(1,0){3}}
\put(1,3){\line(1,0){3}} \put(.5,.3){2} \put(.5,1.3){1}
\put(.5,2.3){0} \put(1.4,3.2){0} \put(2.4,3.2){1} \put(3.4,3.2){2}
% In the grid
\put(1.4,2.3){1} \put(2.4,1.3){1} \put(3.4,.3){1} \put(1.4,1.3){2}
\put(1.4,.3){0} \put(2.4,2.3){0} \put(2.4,.3){2} \put(3.4,1.3){0}
\put(3.4,2.3){2} \put(-.1,3){$B$} \put(.6,3.6){$A$}
\end{picture} \end{center} \vspace{.1 cm}

We begin at $|\psi_3\rangle$, after the states have been initialized.
\[
|\psi_3\rangle = \frac{1}{3}
\left( \begin{array}{ccccccccc}
\omega^2 & 0 & 0 & 0 & 0 & 0 & 0 & 0 & 0\\
0 & \omega & 0 & 0 & 0 & 0 & 0 & 0 & 0\\
0 & 0 & 1 & 0 & 0 & 0 & 0 & 0 & 0\\
0 & 0 & 0 & 1 & 0 & 0 & 0 & 0 & 0\\
0 & 0 & 0 & 0 & \omega^2 & 0 & 0 & 0 & 0\\
0 & 0 & 0 & 0 & 0 & \omega & 0 & 0 & 0\\
0 & 0 & 0 & 0 & 0 & 0 & \omega & 0 & 0\\
0 & 0 & 0 & 0 & 0 & 0 & 0 & 1 & 0\\
0 & 0 & 0 & 0 & 0 & 0 & 0 & 0 & \omega^2
\end{array} \right)
\left( \begin{array}{c}
1 \\
1 \\
1 \\
1 \\
1 \\
1 \\
1 \\
1 \\
1
\end{array} \right) = \frac{1}{3}
\left( \begin{array}{c}
\omega^2 \\
\omega \\
1 \\
1 \\
\omega^2 \\
\omega \\
\omega \\
1 \\
\omega^2
\end{array} \right)
\]

Lemma 1 is used for simplification ($1+\omega+\omega^2=0$):

\[
{\cal{F}}_n^{\otimes 2}|\psi_3\rangle = \frac{1}{9}
\left( \begin{array}{ccccccccc}
1 & 1 & 1 & 1 & 1 & 1 & 1 & 1 & 1\\
1 & \omega & \omega^2 & 1 & \omega & \omega^2 & 1 & \omega & \omega^2\\
1 & \omega^2 & \omega & 1 & \omega^2 & \omega & 1 & \omega^2 & \omega\\
1 & 1 & 1 & \omega & \omega & \omega & \omega^2 & \omega^2 & \omega^2\\
1 & \omega & \omega^2 & \omega & \omega^2 & 1 & \omega^2 & 1 & \omega\\
1 & \omega^2 & \omega & \omega & 1 & \omega^2 & \omega^2 & \omega & 1\\
1 & 1 & 1 & \omega^2 & \omega^2 & \omega^2 & \omega & \omega & \omega\\
1 & \omega & \omega^2 & \omega^2 & 1 & \omega & \omega & \omega^2 & 1\\
1 & \omega^2 & \omega & \omega^2 & \omega & 1 & \omega & 1 &
\omega^2
\end{array} \right) \left( \begin{array}{c}
\omega^2 \\
\omega \\
1 \\
1 \\
\omega^2 \\
\omega \\
\omega \\
1 \\
\omega^2
\end{array} \right) =
\left( \begin{array}{c}
0 \\
0 \\
0 \\
0 \\
0 \\
0 \\
0 \\
\omega^2 \\
0
\end{array} \right)
\]\[
\implies |\psi_4\rangle = \omega^2 |21\rangle,
\]
from which we derive a closed expression for the affine function $f(x_1, x_2)=A_0 \oplus A_1x_1 \oplus A_2x_2$ in $U_f$, save for the constant term, by taking $\{2, 1\}$ as the respective coefficients $A_1$ and $A_2$ (although theoretically, $A_0$ should be $(-2) \bmod 3 = 1$):
\[
f(x_1, x_2)=\mbox{(Constant)}\oplus 2x_1\oplus x_2.
\]
\vspace{5 mm}

\noindent \textbf{Example 2} (Deutsch-Jozsa for a non-affine function)

$U_f$ contains the following balanced function defined on two qutrits $|AB\rangle$:

\vspace{.1 cm} \begin{center} \setlength{\unitlength}{.5 cm}
\begin{picture}(4,4)
\put(1,3){\line(-1,1){.8}} \put(1,0){\line(0,1){3}}
\put(2,0){\line(0,1){3}} \put(3,0){\line(0,1){3}}
\put(4,0){\line(0,1){3}} \put(1,0){\line(1,0){3}}
\put(1,1){\line(1,0){3}} \put(1,2){\line(1,0){3}}
\put(1,3){\line(1,0){3}} \put(.5,.3){2} \put(.5,1.3){1}
\put(.5,2.3){0} \put(1.4,3.2){0} \put(2.4,3.2){1} \put(3.4,3.2){2}
% In the grid
\put(1.4,2.3){0} \put(2.4,1.3){0} \put(3.4,.3){1} \put(1.4,1.3){2}
\put(1.4,.3){1} \put(2.4,2.3){1} \put(2.4,.3){2} \put(3.4,1.3){0}
\put(3.4,2.3){2} \put(-.1,3){$B$} \put(.6,3.6){$A$}
\end{picture} \end{center} \vspace{.1 cm}

Again beginning at $|\psi_3\rangle$:

\[
|\psi_3\rangle = \frac{1}{3}\left( \begin{array}{ccccccccc}
1 & 0 & 0 & 0 & 0 & 0 & 0 & 0 & 0\\
0 & \omega & 0 & 0 & 0 & 0 & 0 & 0 & 0\\
0 & 0 & \omega^2 & 0 & 0 & 0 & 0 & 0 & 0\\
0 & 0 & 0 & \omega^2 & 0 & 0 & 0 & 0 & 0\\
0 & 0 & 0 & 0 & 1 & 0 & 0 & 0 & 0\\
0 & 0 & 0 & 0 & 0 & \omega & 0 & 0 & 0\\
0 & 0 & 0 & 0 & 0 & 0 & \omega & 0 & 0\\
0 & 0 & 0 & 0 & 0 & 0 & 0 & 1 & 0\\
0 & 0 & 0 & 0 & 0 & 0 & 0 & 0 & \omega^2
\end{array} \right)
\left( \begin{array}{c}
1 \\
1 \\
1 \\
1 \\
1 \\
1 \\
1 \\
1 \\
1
\end{array} \right) = \frac{1}{3}
\left( \begin{array}{c}
1 \\
\omega \\
\omega^2 \\
\omega^2 \\
1 \\
\omega \\
\omega \\
1 \\
\omega^2
\end{array} \right),
\]\[
{\cal{F}}_n^{\otimes 2}|\psi_3\rangle = \frac{1}{9}
\left( \begin{array}{ccccccccc}
1 & 1 & 1 & 1 & 1 & 1 & 1 & 1 & 1\\
1 & \omega & \omega^2 & 1 & \omega & \omega^2 & 1 & \omega & \omega^2\\
1 & \omega^2 & \omega & 1 & \omega^2 & \omega & 1 & \omega^2 & \omega\\
1 & 1 & 1 & \omega & \omega & \omega & \omega^2 & \omega^2 & \omega^2\\
1 & \omega & \omega^2 & \omega & \omega^2 & 1 & \omega^2 & 1 & \omega\\
1 & \omega^2 & \omega & \omega & 1 & \omega^2 & \omega^2 & \omega & 1\\
1 & 1 & 1 & \omega^2 & \omega^2 & \omega^2 & \omega & \omega & \omega\\
1 & \omega & \omega^2 & \omega^2 & 1 & \omega & \omega & \omega^2 & 1\\
1 & \omega^2 & \omega & \omega^2 & \omega & 1 & \omega & 1 &
\omega^2
\end{array} \right) \left( \begin{array}{c}
1 \\
\omega \\
\omega^2 \\
\omega^2 \\
1 \\
\omega \\
\omega \\
1 \\
\omega^2
\end{array} \right) =
\left( \begin{array}{c}
0 \\
\omega/3 \\
1/3+\omega^2/3 \\
0 \\
1/3 \\
2/3 \\
0 \\
\omega^2/3 \\
1/3+\omega/3
\end{array} \right),
\]\[
|\psi_4\rangle = \frac{\omega}{3}|01\rangle + \frac{1+\omega^2}{3}|02\rangle + \frac{1}{3}|11\rangle + \frac{2}{3}|12\rangle + \frac{\omega^2}{3}|21\rangle + \frac{1+\omega}{3}|22\rangle.
\]
The basis state $|12\rangle$ hence has a $4/9$ probability of measurement, with all others having $1/9$ probability.  To determine that this balanced function is non-affine, enough measurements of the $x$-register are required so as to obtain different states at output.  Furthermore, observe that if the function is not affine, there is still a relatively high probability of measuring the output state of the algorithm that would be associated with a similar affine function.  For example, compare the Marquand chart of the non-affine function in $U_f$ above with that of the affine function $f(x_1, x_2)=x_1\oplus 2x_2$ associated with the output state $|12\rangle$:

\vspace{.1 cm} \begin{center} \setlength{\unitlength}{.5 cm}
\begin{picture}(4,4)
\put(1,3){\line(-1,1){.8}} \put(1,0){\line(0,1){3}}
\put(2,0){\line(0,1){3}} \put(3,0){\line(0,1){3}}
\put(4,0){\line(0,1){3}} \put(1,0){\line(1,0){3}}
\put(1,1){\line(1,0){3}} \put(1,2){\line(1,0){3}}
\put(1,3){\line(1,0){3}} \put(.5,.3){2} \put(.5,1.3){1}
\put(.5,2.3){0} \put(1.4,3.2){0} \put(2.4,3.2){1} \put(3.4,3.2){2}
% In the grid
\put(1.4,2.3){0} \put(2.4,1.3){0} \put(3.4,.3){0} \put(1.4,1.3){2}
\put(1.4,.3){1} \put(2.4,2.3){1} \put(2.4,.3){2} \put(3.4,1.3){1}
\put(3.4,2.3){2} \put(-.1,3){$x_2$} \put(.6,3.6){$x_1$}
\end{picture} \end{center} \vspace{.1 cm}

Because an affine multi-valued function in our context is defined in terms of the modulo-additive operator $\oplus$, any arbitrary function is affine iff it satisfies the \emph{cyclic group property} (this is easy to determine when the outputs are plotted in a Marquand chart as above, which requires that rows and columns be successive cyclic shifts of each other), and the state after $U_f$ is factorable - e.g., entanglement does not occur in the Deutsch-Jozsa circuit.

Our extended Deutsch-Jozsa algorithm requires only one measurement to deterministically distinguish an expression for an affine function of radix $n$ and $r$ inputs up to the accuracy of a constant, given that it is affine.

\section{The Grover Algorithm in Radix \emph{n}}

The quantum Fourier transform could also be applied to Grover's quantum search algorithm, which, in a multi-valued context, would reduce the number of qudits necessary in the memory register for large search spaces.  Specifically, while a search space of $N$ elements in the binary Grover algorithm would require $\lceil\lg N\rceil$ qubits to implement, a radix-$n$ circuit would require $\lceil\log_n N\rceil$ qudits.  This reduces the number of qudits by a factor of about $\lg n$.
\vspace{5 mm}

Take $|\psi\rangle=({\cal{F}}_n|0\rangle)^{\otimes r}$ as the vector representing the $r$-qudit search space of $N$ elements in an equal superposition, $N=n^r$, and $|i_0\rangle$ as the searched state.  The action of the oracle $U_f|xy\rangle = |x\rangle|y \oplus f(x)\rangle$ gives us reason to redefine the (now ill-termed) inversion-about-the-average, or diffusion, operator as
\[
D = (1-\omega^{-1})|\psi\rangle\langle\psi|-I
\]
and the selective phase rotation operator $U_{|i_0\rangle}$ as
\[
I-(1-\omega^{-1})|i_0\rangle\langle i_0|,
\]
if only the action on the $x$-register is considered.  It is easily verified that both of these operators are unitary.  The expression $(1-\omega^{-1})$ appears because $U_f$, adding $f(x)$ to the basis states $y$ modulo $n$, accordingly rotates the phase of the searched state by $-2\pi/n$.  $U_{|i_0\rangle}$ can be constructed from $U_f$ by transforming the $y$-register, initialized to $|1\rangle$, via QFT to the superposed state $\frac{1}{\sqrt{n}}\sum_{k=0}^{n-1}\omega^k |k\rangle$, as shown below \cite{J-on-G}.

\begin{figure}[h]
\hspace{.5 cm}\centerline{ \Qcircuit @C=2em @R=2.5em {\lstick{\ket{\psi}} & \multigate{1}{\hspace{.6 cm} U_f \hspace{.5 cm}} & \qw & \lstick{U_{|i_0\rangle}\ket{\psi} \hspace{-1 cm}}\\
\lstick{\displaystyle{\frac{1}{\sqrt{n}}\sum_{k=0}^{n-1}\omega^k
|k\rangle}} & \ghost{\hspace{.6 cm} U_f \hspace{.5 cm}} & \qw &
\lstick{\displaystyle{\frac{1}{\sqrt{n}}\sum_{k=0}^{n-1}\omega^k
|k\rangle} \hspace{-1.7 cm}}}} \vspace{1 mm}
\caption{Implementation of $U_{|i_0\rangle}$}
\end{figure}
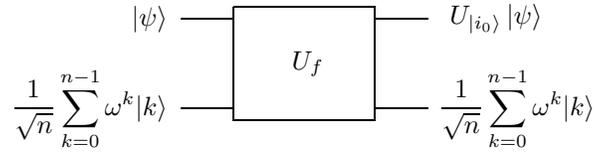

\noindent Again, if the $y$-register is ignored for computational purposes, then the oracle takes the same form as $U_{|i_0\rangle}$:
\[
U_f = \sum_{x=0}^{{n^r}-1} \omega^{-f(x)} |x\rangle\langle x|,
\]
where $f(x)=1$ if $x=i_0$ and $f(x)=0$ otherwise, thereby marking the searched state through phase rotation.

Hence, the entire Grover iteration can be expressed as the unitary operator
\[
G = D U_{|i_0\rangle}=(1-\omega^{-1})(|\psi\rangle\langle\psi|+|i_0\rangle\langle i_0|)-\frac{(1-\omega^{-1})^2}{\sqrt{N}}|\psi\rangle\langle i_0|-I.
\]
For practical implementation, the multi-valued $D$ operator can also be decomposed into a QFT, a selective phase rotation of the state $|0\rangle$ by $-2\pi/n$, and an inverse QFT:
\[
(1-\omega^{-1})|\psi\rangle\langle\psi|-I = {\cal{F}}_n^{\otimes r}[(1-\omega^{-1})|0\rangle\langle0| - I]({\cal{F}}_n^\dagger)^{\otimes r} = -{\cal{F}}_n^{\otimes r}U_{|0\rangle}({\cal{F}}_n^\dagger)^{\otimes r}
\]

Since the multi-valued Grover algorithm deals with rotations through complex angles in $n$-dimensional Hilbert space whereas the binary algorithm deals only with rotations through real angles in a real subspace of $2^r$-dimensional Hilbert space, the matter of precisely how many times to iterate is more complicated.  Although each Grover iteration does rotate $|\psi\rangle$ through a complex angle $\theta$ where $\cos\theta = \langle\psi|G^{k+1}G^k|\psi\rangle$ after $k$ iterations, the interpretation of such a rotation as increasing the component of $|\psi\rangle$ in the direction of $|i_0\rangle$ (\cite{Lavor}, \cite{J-on-G}) is not as straightforward in higher radices. It is easily seen, though, that as $n$ becomes very large, the number of necessary iterations increases since the $(1-\omega^{-1})$ factor in $G$ approaches 0.  Thus we find a tradeoff between time and space complexity when going to multi-valued logic.
\vspace{5 mm}

We now argue that our multi-valued extension of the Grover algorithm still converges to the desired state in $O(\sqrt{N})$ steps.  In the subspace of $\mathcal{H}_n^{\otimes r}$ spanned by (the non-orthogonal basis) $\{|i_0\rangle, |\psi\rangle\}$, we have the eigendecomposition
\[
G = \left(\begin{array}{cc} -\omega^{-1} & \frac{1 - \omega^{-1}}{\sqrt{N}} \\ \frac{\omega^{-1}(1 - \omega^{-1})}{\sqrt{N}} & -\omega^{-1} - \frac{(1 - \omega^{-1})^2}{N} \end{array}\right) = M\left(\begin{array}{cc} \lambda_+ & 0 \\ 0 & \lambda_- \end{array}\right)M^{-1} % = M^{-1}\left(\begin{array}{cc} \lambda_+ & 0 \\ 0 & \lambda_- \end{array}\right)M
\]
with eigenvalues
\[
\lambda_{\pm} = -\omega^{-1} - \frac{(1 - \omega^{-1})^2}{2N} \pm \frac{1 - \omega^{-1}}{2}\sqrt{\frac{4\omega^{-1}}{N} + \frac{(1 - \omega^{-1})^2}{N^2}}
\]
and
\[
M = \left(\begin{array}{cc} 1 & 1 \\ -\frac{1 - \omega^{-1}}{2\sqrt{N}} + \sqrt{\omega^{-1} + \frac{(1-\omega^{-1})^2}{4N}} & -\frac{1 - \omega^{-1}}{2\sqrt{N}} - \sqrt{\omega^{-1} + \frac{(1-\omega^{-1})^2}{4N}} \end{array}\right) \equiv \left(\begin{array}{cc} 1 & 1 \\ x & y \end{array}\right). % M = \left(\begin{array}{cc} 1 & 1 \\ \frac{1 + \omega^{-1}}{2} - \frac{1 - \omega^{-1}}{2\sqrt{N}} & -\frac{1 + \omega^{-1}}{2} - \frac{1 - \omega^{-1}}{2\sqrt{N}} \end{array}\right) \equiv \left(\begin{array}{cc} 1 & 1 \\ x & y \end{array}\right).
\]
Suppose $N\gg 1$.  We calculate the amplitude of $|i_0\rangle$ after $k$ iterations.  Keeping only terms of lowest order in $1/N$, namely $1/2$, and assuming $k\ll N$ over the course of the evolution that we consider (this latter assumption will be justified at the end of the computation), we have
\begin{eqnarray*}
\langle i_0|G^k|\psi\rangle &=& (\begin{array}{cc} 1 & 1/\sqrt{N} \end{array}) M \left(\begin{array}{cc} \lambda_+^k & 0 \\ 0 & \lambda_-^k \end{array}\right) M^{-1} \left(\begin{array}{c} 0 \\ 1 \end{array}\right) \\ % \left(\begin{array}{cc} 1 & 1/\sqrt{N} \end{array}\right) M^{-1}\left(\begin{array}{cc} -\omega^{-1} + \frac{\omega^{-2}(1 - \omega^{-1})}{\sqrt{N}} & 0 \\ 0 & -\omega^{-1} - \frac{\omega^{-2}(1 - \omega^{-1})}{\sqrt{N}} \end{array}\right)^k M \left(\begin{array}{c} 0 \\ 1 \end{array}\right) \\
&=& \frac{1}{y-x}\left[\left(1 + \frac{y}{\sqrt{N}}\right)\left(-\omega^{-1} - \frac{\omega^{-1/2}(1 - \omega^{-1})}{\sqrt{N}}\right)^k - \right. \\
&& \hspace{1.13 cm} \left.\left(1 + \frac{x}{\sqrt{N}}\right)\left(-\omega^{-1} + \frac{\omega^{-1/2}(1 - \omega^{-1})}{\sqrt{N}}\right)^k\right] \\ % \frac{1}{y - x}\left(y(\lambda_+^k - \lambda_-^k) + \frac{y}{\sqrt{N}}\lambda_-^k - \frac{x}{\sqrt{N}}\lambda_+^k\right) \\
&=& \frac{1}{y-x}\left[-2k(-\omega^{-1})^{k-1}\frac{\omega^{-1/2}(1-\omega^{-1})}{\sqrt{N}} + \frac{y}{\sqrt{N}}(-\omega^{-1})^k - \frac{x}{\sqrt{N}}(-\omega^{-1})^k\right] \\ % -\frac{1}{1 + \omega^{-1}}\left(2yk(-\omega^{-1})^{k-1}\frac{\omega^{-2}(1 - \omega^{-1})}{\sqrt{N}} + \frac{y}{\sqrt{N}}(-\omega^{-1})^k - \frac{x}{\sqrt{N}}(-\omega^{-1})^k\right) \\
&=& \frac{(-\omega^{-1})^{k-1}}{\sqrt{N}}\left[k(1 - \omega^{-1}) - \omega^{-1}\right]. % \frac{(-\omega^{-1})^{k-1}}{\sqrt{N}}(k\omega^{-2}(1 - \omega^{-1}) - \omega^{-1}).
\end{eqnarray*}
All the terms in $\omega$ are of bounded and small magnitude; hence $k$ comes to dominate fairly quickly, and we see that the amplitude of $|i_0\rangle$ in the state of the register grows linearly with $k$.  Furthermore, since the coefficient of $k$ is on the order of $1/\sqrt{N}$, $O(\sqrt{N})$ steps are required to increase the magnitude of the amplitude of the target state beyond any given threshold, as desired, consistent with the assumption that $k\ll N$ for the entire evolution.

\section{Applications and Conclusion}

We have demonstrated that the multi-valued Deutsch-Jozsa algorithm is able to test for the constancy or affinity of a multi-valued logical function in a single step, preserving exponential speedup over classical algorithms.  Furthermore, the multi-valued Grover algorithm allows for a reduction in memory register size by using fewer qudits to represent more information states, reduces the number of ``wasted'' information states by searching unsorted databases of close to arbitrary size, and still affords a quadratic speedup over classical algorithms.

While the original Deutsch-Jozsa algorithm is mainly of theoretical interest, its multi-valued extension could potentially find application in image processing to distinguish between maps of colored patterns represented by affine functions in a Marquand chart.  A salient feature of texture recognition is identifying and comparing predictably repeating patterned images.  Such images could be easily encoded by maps of affine functions with each pixel represented by an output value, and each color represented by a different radix-$n$ digit, $n$ corresponding to the number of colors in the image.  Since we have shown that the multi-valued Deutsch-Jozsa algorithm gives a unique deterministic output for any class of affine function (i.e., differing in only the constant term) in $O(1)$ time, it presents a very efficient method of characterizing a texture image by a unique ``signature'' (the output state) without using the substantial processing power required by classical means.  Since textures repeat, the constant term would be insignificant since it serves only to permute the entries in the function map.  The image would only have to be encoded in the oracle, which would require a quantum computer with programmable gate arrays, and could be realized with prospective technology.

Also, Grover's algorithm is well-known for its pertinence to a wide range of NP problems for which it could be adapted by modifying the oracle architecture.  Several practical problems that could be solved include graph coloring, the traveling salesman problem (to which the Grover algorithm could be applied to verify whether any path has a cost less than a given threshold), and the general NP-complete Boolean satisfiability (SAT) problem.  For example, in searching for a valid coloring of a planar map (where no two adjacent regions are of the same color) from the set of all possible colorings, any coloring of the corresponding graph with $s$ vertices could be encoded in a string of $s$ radix-4 digits; hence, the 4-valued Grover algorithm would be most efficient in terms of memory register size.  Grover's algorithm is very versatile in that it can be modified for any general class of search problem; given a memory register of all possible inputs, the oracle can check each element against a Boolean or multi-valued logical function constructed from a certain number of constraints.  Hence, most applications of Grover's algorithm are based on the reducibility of many important problems to SAT.  The SAT problem could potentially find even more application when reformulated for multi-valued logic to allow for different numbers of parameters, using modulo-$n$ addition and multiplication as natural generalizations of Boolean XOR and AND.

The fact that the number of iterations needed to effectively implement Grover's algorithm increases with radix size leads to the question of the ``optimal'' radix for Grover's algorithm.  This is a topic for further research.  However, from numerical analysis in MATLAB, we conjecture that ternary logic is optimal for quantum searching in that it requires the smallest number of iterations for a search space of any given size.  This may be related to its closeness to base $e$, the most ``natural'' base for computing.

\section*{Acknowledgements}

The author gratefully acknowledges Professor Marek Perkowski of the Portland State University Department of Electrical and Computer Engineering for his guidance and Jacob Biamonte of D-Wave Systems, as well as the members of the Portland Quantum Logic Group for their support.

\section*{Appendix}

In the multi-valued Deutsch-Jozsa algorithm, for radices higher than two in which $U_f$ is merely $r+1$ wires in parallel (implementing the constant function $f:\{0,1\}^r \rightarrow 0$), the initialized state $|0\rangle^{\otimes r}|1\rangle$ is mapped to a different output $|0\rangle^{\otimes r}|n-1\rangle$.  Practically, this makes no difference because the output of the $y$-register is discarded.  However, Theorem 3, and the related characterization of ${\cal{F}}_n^2$, make clear why this is so.

\begin{fourth}
Four iterations of the QFT gives an identity mapping (the binary Hadamard gate is a special case that is also self-inverse).
\end{fourth}
\emph{Proof.} Let $[{\cal{F}}_n]_{pq}$ denote the $p,q$ entry of ${\cal{F}}_n$.  Row $p$ and column $q$ of ${\cal{F}}_n$ are given by
\[
\langle p|=\frac{1}{\sqrt{n}} \sum_{k=0}^{n-1} \omega^{k(p-1)}\langle k| \mbox{ and } |q\rangle=\frac{1}{\sqrt{n}} \sum_{k=0}^{n-1} \omega^{k(q-1)}|k\rangle,
\]
respectively.  We have $[{\cal{F}}_n^2]_{pq} = \langle p | q \rangle = \frac{1}{n} \sum_{k=0}^{n-1}\omega^{k(p+q-2)}$; $p+q-2$ is some integer $\alpha$ that is nonzero when $(p+q)\bmod n \neq 2$, so by Lemma 1, $[{\cal{F}}_n^2]_{pq}=1$ if $(p+q)\bmod n=2$ and $[{\cal{F}}_n^2]_{pq}=0$ otherwise.  Then all such indices $p,q \in \{1, 2, \ldots, n\}$ for which $[{\cal{F}}_n^2]_{pq}=1$ satisfy $p+q=C_0n+2$ for some $C_0 \in \mathbf{Z}_n$.  $C_0$ is further restricted to $\{0,1\}$ because any larger values of $C_0n+2$ exceed the maximum value of $p+q$, or $2n$.  Therefore, either $p+q=2$ or $p+q=n+2$, giving the solution sets
\[
p=q=1 \mbox{ or } \{p,q\}=\{n-m+1,m+1\} \mbox{ for } m \in \{1,2,\ldots, n-1\},
\]
which correspond to the permutation matrix
\[
{\cal{F}}_n^2 = \sum_{m=0}^{n-1} |n-m\rangle\langle m|,\mbox{ } |n\rangle \mbox{ taken to mean } |0\rangle.
\]
Thus
\begin{eqnarray*}
{\cal{F}}_n^3 &=&
\frac{1}{\sqrt{n}}\sum_{j=0}^{n-1}\sum_{k=0}^{n-1}\sum_{m=0}^{n-1}e^{i2\pi
jk/n}|n-m\rangle\langle m | j\rangle\langle k| \\
&=& \frac{1}{\sqrt{n}}\sum_{j=0}^{n-1}\sum_{k=0}^{n-1}e^{i2\pi
jk/n}|n-j\rangle\langle k| \\
&=& \frac{1}{\sqrt{n}}\sum_{j=0}^{n-1}\sum_{k=0}^{n-1}e^{-i2\pi
jk/n}|j\rangle\langle k|\\
&=& {\cal{F}}_n^\dag
\end{eqnarray*}
by our previous arguments, and given that ${\cal{F}}_n$ is unitary,
\[
{\cal{F}}_n^4 = {\cal{F}}_n{\cal{F}}_n^\dag = I_n
\]
is immediate.  This is a property that the QFT shares with the continuous Fourier transform.

\end{document}